\documentclass[aps,pra,twocolumn,amsmath,amssymb,nofootinbib,superscriptaddress]{revtex4-1}

\usepackage{times}
\usepackage[pdftex]{graphicx}
\usepackage{dcolumn}
\usepackage{bm}
\usepackage{amsmath}
\usepackage{indentfirst}
\usepackage{float}
\usepackage[colorlinks]{hyperref}
\usepackage[dvipsnames]{xcolor}
\usepackage{xcolor}
\usepackage{subfigure}
\usepackage{algorithm}
\usepackage{algorithmicx}
\usepackage{algpseudocode}
\usepackage{amsmath}
\usepackage{verbatim}
\usepackage{makecell}
\usepackage[normalem]{ulem}

\newcommand{\Eloc}{E_{{\rm loc}}}

\newcommand{\hnu}{Key Laboratory of Low-Dimensional Quantum Structures and Quantum Control of Ministry of Education, Department of Physics and Synergetic Innovation Center for Quantum Effects and Applications, Hunan Normal University, Changsha 410081, China
}
\newcommand{\cas}{Institute of Computing Technology, Chinese Academy of Sciences, Beijing, China}

\newcommand{\sutd}{Science, Mathematics and Technology Cluster, Singapore University of Technology and Design, 8 Somapah Road, 487372 Singapore} 
\newcommand{\sutdepd}{EPD Pillar, Singapore University of Technology and Design, 8 Somapah Road, 487372 Singapore} 

\newcommand{\ustc}{University of Science and Technology of China, Hefei, China}

\newcommand{\sicnu}{College of Physics and Electronic Engineering, and Center for Computational Sciences, Sichuan Normal University, Chengdu 610068, China}

\begin{document}

\title{A real neural network state for quantum chemistry}

\author{Yangjun Wu}
\affiliation{\cas}

\author{Xiansong Xu} 
\affiliation{\sutd}
\affiliation{\sicnu}

\author{Dario Poletti}
\affiliation{\sutd} 
\affiliation{\sutdepd} 

\author{Yi Fan}
\affiliation{\ustc}

\author{Chu Guo}
\email{guochu604b@gmail.com}
\affiliation{Henan Key Laboratory of Quantum Information and Cryptography, Zhengzhou,
Henan 450000, China}
\affiliation{\hnu}

\author{Honghui Shang}
\email{shanghonghui@ict.ac.cn}
\affiliation{\cas}



\begin{abstract}
The restricted Boltzmann machine (RBM) has been successfully applied to solve the many-electron Schr$\ddot{\text{o}}$dinger equation. In this work we propose a single-layer fully connected neural network adapted from RBM and apply it to study ab initio quantum chemistry problems. Our contribution is two-fold: 1) our neural network only uses real numbers to represent the real electronic wave function, while we obtain comparable precision to RBM for various prototypical molecules; 2) we show that the knowledge of the Hartree-Fock reference state can be used to systematically accelerate the convergence of the variational Monte Carlo algorithm as well as to increase the precision of the final energy.
\end{abstract}

\maketitle

\section{Introduction}\label{s:intro}
Ab initio electronic structure calculations based on quantum-chemical approaches (Hartree–Fock theory and post-Hartree–Fock methods) have been successfully applied in molecular systems~\cite{molecular_electronic_structure_theory}. 
For strongly correlated many-electron systems, the exponentially growing Hilbert space size limits the application scale of most numerical algorithms. For example, the full configuration interaction (FCI) which takes the whole Hilbert space into account, is currently limited within around $24$ orbitals and $24$ electrons~\cite{VogiatzisJong2017}. The density matrix renormalization group (DMRG) algorithm~\cite{White1992,White1993} has been used to solve larger chemical systems of several tens of electrons~\cite{BrabecVeis2020,LarssonChan2022}, however it is essentially limited by the expressive power of its underlying variational ansatz: the matrix product state (MPS) which is a special instance of the one-dimensional tensor network state~\cite{GarciaCirac2007}, therefore DMRG could also be extremely difficult to approach even larger systems. 
The coupled cluster (CC)~\cite{CCSD-Purvis1982AFC,Cizek1966} method expresses the exact wave function in terms of an exponential form of a variational wave function ansatz, and higher level of accuracy can be obtained by considering electronic excitations up to doublets in CCSD or triplets in CCSD(T). In practice, it is often accurate with a durable computational cost, thus considered as the “gold standard” in electronic structure calculations. However, the accuracy of the CC method is only restricted in studying weakly correlated systems~\cite{CoesterKummel1960}. 
The multi-configuration self-consistent field (MCSCF)~\cite{MCSCF,MCSCF1-KNOWLES1985259,MCSCF2-Jensen1994} method is crucial for describing molecular systems containing nearly degenerate orbitals. It introduces a small number of (active) orbitals, then the configuration interaction coefficients and the orbital coefficients are optimized to minimize the total energy  of the MCSCF state. It has been applied to systems with around 50 active orbitals~\cite{Sun_2020}, but they are still limited by the exponential complexity that grows with the system size.

In recent years the variational Monte Carlo (VMC) method in combination with a neural network ansatz for the underlying quantum state (wave function)~\cite{CarleoTroyer2017}, referred to as the neural network quantum states (NNQS), has been demonstrated to be a scalable and accurate tool for many-spin systems~\cite{ChooCarleo2019,SchmittHeyl2020,YuanDeng2021} and many-fermion systems~\cite{MorenoStokes2022}. 
NNQS allow very flexible choices of the neural network ansatz, and with an appropriate variational ansatz, it could often achieve comparable or higher accuracy compared to existing methods.
NNQS has also been applied to solve ab-initio quantum chemistry systems in real space with up to $30$ electrons~\cite{PauliNet,FermiNet,HumeniukWang2022}, as well as in discrete basis after second quantization~\cite{ChooCarleo2020,BarrettLvovsky2022,ZhaoVeerapaneni2022}. 
Up to now various neural networks have been used, such as the restricted Boltzmann machine (RBM)~\cite{CarleoTroyer2017}, convolutional neural network~\cite{ChooCarleo2019}, recurrent neural networks~\cite{CarleoWu2022} and variational auto-encoder~\cite{ZhaoVeerapaneni2022}. In all those neural networks, the RBM is a very special instance in that: 1) it has a very simple structure which contains only a fully connected dense layer plus a nonlinear activation; 2) with such a simple structure, RBM can be more expressive than MPS~\cite{SharirCarleo2022}, in fact it is equivalent to certain two-dimensional tensor network states~\cite{GlasserCirac2018}, and can even represent certain quantum state with volume-law entanglement~\cite{DengSarma2017}. In practice RBM achieves comparable accuracy to other more sophisticated neural networks for complicated applications such as frustrated many-spin systems~\cite{NomuraImada2021,LiangWei2022}.

For the ground state of molecular systems, the wave function is real.
However, if one uses a real RBM as the variational ansatz for the wave function, then all the amplitudes of the wave function will be positive, which means that it may be good for ferromagnetic states but will be completely wrong for anti-ferromagnetic states. Therefore even for real wave functions one would have to use complex RBMs or two RBMs~\cite{TorlaiCarleo2018} in general. In this work we propose a neural network with real numbers which is slightly modified from the RBM such that its output can be both positive and negative, and use it as the neural network ansatz to solve quantum chemistry problems. To accelerate convergence of the VMC iterations, we explicitly use the Hartree-Fock reference state as the starting point for the Monte Carlo sampling after a number of VMC iterations such that the wave function ansatz has become sufficiently close to the ground state. We show that this technique can generally improve the convergence and the precision of the final result, even when using other neural networks. Our paper is organized as follows. In Sec.~\ref{sec:methods} we present our neural network ansatz. In Sec.~\ref{sec:result} we present our numerical results demonstrating the effectiveness of our neural network ansatz and the technique of initializing the Monte Carlo sampling with the Hartree-Fock reference state. We conclude in Sec.~\ref{sec:summary}.







\section{Methods}\label{sec:methods}

\subsection{Real neural network ansatz}

\begin{figure}
\centering
\includegraphics[width=\columnwidth]{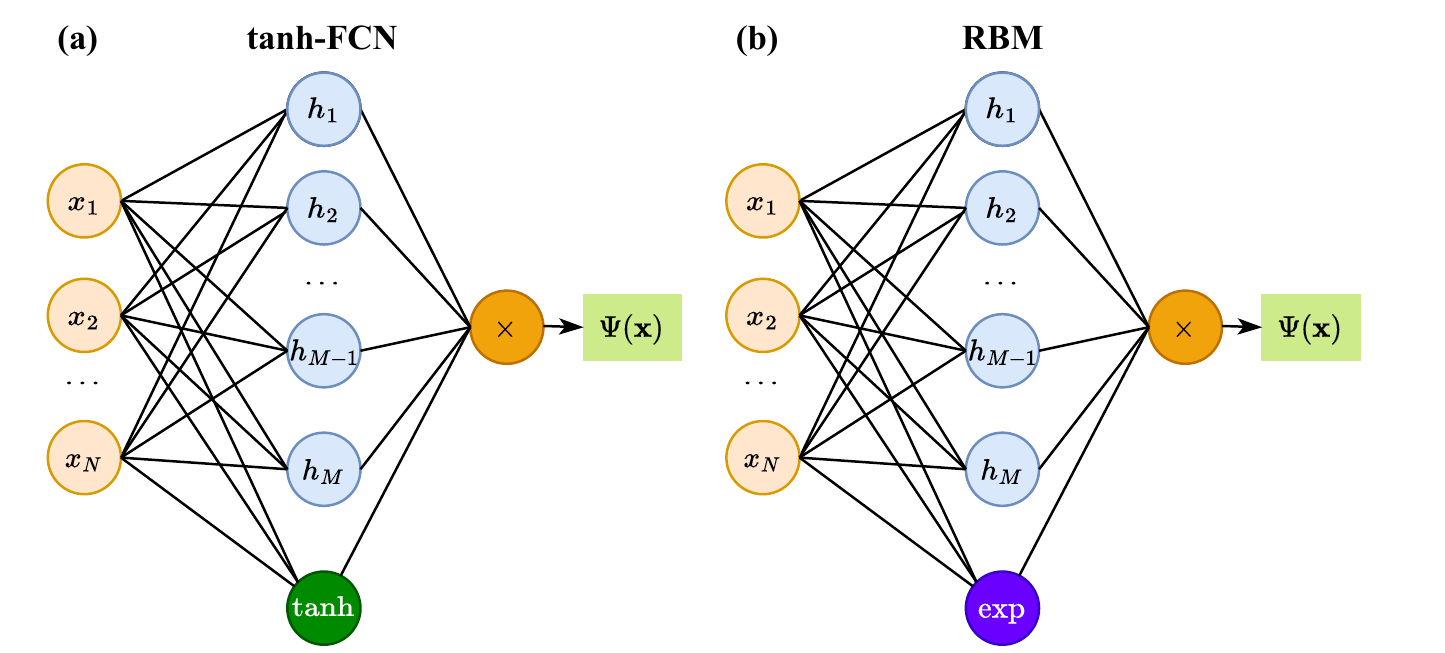}
\caption{The architectures for (a) our tanh-FCN and (b) RBM. The major difference is that we use hyperbolic tangent as the activation function such that  tanh-FCN could output both positive and negative numbers even if it only uses real numbers.}
\label{fig:fcn}
\end{figure}

Before we introduce our model we first briefly review the RBM used in NNQS. For a classical many-spin system, one could embed the system into a larger one consisting of visible spins (corresponding to the system) and hidden spins with the total (classical) Hamiltonian
\begin{align}
\mathcal{H} = \sum_{j=1}^{N_v} a_j x_j + \sum_{i=1}^{N_h} b_i h_i + \sum_{i, j} W_{ij}h_ix_j,
\end{align}
where $x_j$ represents the visible spin and $h_i$ the hidden spin. $N_v$ and $N_h$ are the number of visible and hidden spins respectively. The coefficients $\theta =\{a, b, W\} $ are variational parameters of the Hamiltonian. Since there is no coupling between the hidden spins, one could explicitly integrate them out and get the partition function of the system $\mathcal{Z}$ as
\begin{align}
\mathcal{Z} = \sum_{\bold{x}} p(\bold{x}),
\end{align}
with $\bold{x} = \{x_1, x_2, \dots, x_{N_v}\}$ a particular configuration and $p(\bold{x})$ the unnormalized probability (in case of real coefficients) of $\bold{x}$, which can be explicitly written as
\begin{align}\label{eq:rbm_prob}
p(\bold{x}) &= \sum_{\bold{h}} e^{\mathcal{H}} \nonumber \\ 
&= e^{\sum_{j=1}^{N_v} a_j x_j} \times \prod_{i=1}^{N_h} 2\cosh(b_i + \sum_{j=1}^{N_v}W_{ij}x_j ).
\end{align}
When using RBM as a variational ansatz for the wave function of a quantum many-spin system, $p(\bold{x})$ is interpreted as the amplitude (instead of the probability) of the configuration $\bold{x}$.
Eq.(\ref{eq:rbm_prob}) can be seen as a single-layer fully connected neural work which accepts a configuration (a vector of integers) as input and outputs a scalar. For real coefficients, 
the output will always be positive by definition, therefore one generally has to use complex coefficients even for real wave functions. In this work, we slightly change Eq.(\ref{eq:rbm_prob}) as follows so as to be able to output any real numbers with a real neural network:
\begin{align}\label{eq:fcn_prob}
p(\bold{x}) = \tanh(\sum_{j=1}^{N_v} a_j x_j) \times \prod_{i=1}^{N_h} 2\cosh(b_i + \sum_{j=1}^{N_v}W_{ij}x_j ).
\end{align}
In the following we will write $p(\bold{x})$ as $\Psi_{\theta}(\bold{x})$ to stress its dependence on the variational parameters and that it is interpreted as a wave function instead of a probability distribution, we will also refer to our neural network in Eq.(\ref{eq:fcn_prob}) as tanh-FCN since it contains a fully connected layer followed by hyperbolic tangent as the activation function. The difference between RBM and tanh-FCN is demonstrated in Fig.~\ref{fig:fcn}.

\subsection{Variational Monte Carlo}
The electronic Hamiltonian $\hat{H}^e$ of a chemical system can be written in a second-quantized formulation:
	\begin{align}
	    \hat{H}^e=\sum_{p,q}{h_{q}^{p}a_{p}^{\dagger}a_{q}} +\frac{1}{2}{\sum_{\substack{p,q\\r,s}}}{g_{rs}^{pq}a_{p}^{\dagger}a_{q}^{\dagger}a_{r}a_{s}}
				\label{eq:ham-pbc}
	\end{align}
	where $h_{q}^{p}$ and $g_{rs}^{pq}$ are one- and two-electron integrals in molecular orbital basis, ${a_{p}^{\dagger}}$ and $a_{q}$ in the Hamiltonian are the creation and annihilation operators. To treat the fermionic systems, we first use the Jordan-Wigner transformation to map the electronic Hamiltonian to a sum of Pauli operators, following Ref.~\cite{ChooCarleo2020}, and then use our tanh-FCN in Eq.(\ref{eq:fcn_prob}) as the ansatz for the resulting many-spin system.
The resulting spin Hamiltonian $\hat{H}$ can generally be written in the following form
\begin{align}
\hat{H}=\sum_i c_i \prod_{j=1}^{N} \sigma_j^{v_{i,j}},
\end{align}
where $N=N_v$ is the number of spins, $c_i$ is a real coefficient and $\sigma_j^{v_{i,j}}$ is a single spin Pauli operator acting on the $j$-th spin ($v_{i,j}\in \{0,1,2,3\}$ and $\sigma^0=I$, $\sigma^1=\sigma^x$, $\sigma^2=\sigma^y$, $\sigma^3=\sigma^z$).

Given the wave function ansatz $\Psi_{\theta}(\bold{x})$, the corresponding energy can be computed as
\begin{align}\label{eq:energy}
E(\theta)=\frac{\langle\Psi_\theta|\hat{H}|\Psi_\theta \rangle}{\langle\Psi_\theta|\Psi_\theta \rangle}
=\frac{\sum_{\bold{x}} \Eloc(\bold{x})\left|\Psi_\theta(\bold{x})\right|^2}{\sum_{\bold{y}}\left|\Psi_\theta(\bold{y})\right|^2},
\end{align}
where the ``local energy'' $\Eloc(\bold{x})$ for a configuration $\bold{x}$ is defined as
\begin{align}
\Eloc(\bold{x})=\sum_{\bold{x}'} \frac{\Psi_\theta(\bold{x}')}{\Psi_\theta(\bold{x})} H_{\bold{x}' \bold{x}},
\end{align}
with $H_{\bold{x}' \bold{x}} = \langle \bold{x}' \vert \hat{H}\vert \bold{x}\rangle $. The VMC algorithm evaluates Eq.(\ref{eq:energy}) approximately using Monte Carlo sampling, namely
\begin{align}\label{eq:energy_sample}
\tilde{E}(\theta) = \langle \Eloc\rangle,
\end{align}
where the average is over a set of samples $\{\bold{x}^1, \bold{x}^2, \dots, \bold{x}^{N_s}\}$ ($N_s$ is the total number of samples), generated from the probability distribution $|\Psi_{\theta}(\bold{x})|^2$. $\tilde{E}(\theta)$ will converge to $E(\theta)$ if $N_s$ is large enough. In this work we use the Metropolis-Hastings sampling algorithm to generate samples~\cite{Hastings1970}. A configuration $\bold{x}$ is updated using the SWAP operation between nearest-neighbour pairs of spins to preserve the electron-number conservation. We also use the natural gradient of Eq.(\ref{eq:energy_sample}) for the stochastic gradient descent algorithm in VMC, namely the parameters are updated as
\begin{align}\label{eq:gd}
\theta^{k+1} = \theta^k - \alpha S^{-1} F,
\end{align}
where $k$ is the number of iterations, $\alpha$ is the learning rate ($\alpha$ is dependent on $k$ in general), $S$ is the stochastic reconfiguration matrix~\cite{SorellaCapriotti2000,SorellaRocca2007} and $F$ is the gradient of Eq.(\ref{eq:energy_sample}). Concretely, $S$ and $F$ are computed by
\begin{align}
S_{ij}(k)=\langle O_i^*O_j \rangle - \langle O_i^* \rangle \langle O_j \rangle,
\end{align}
and 
\begin{align}
F_i(k) = \langle \Eloc O_i^*\rangle - \langle \Eloc \rangle \langle O_i^* \rangle
\end{align}
respectively, with $O_i(\bold{x})$ defined as
\begin{align}
O_i(\bold{x})=\frac{1}{\Psi_\theta(\bold{x})}\frac{\partial \Psi_\theta(\bold{x})}{\partial \theta_i}.
\end{align}
In general $S$ can be non-invertible, and a simple regularization is to add a small shift to the diagonals of $S$, namely using $S^{reg}=S+\epsilon I$ instead of $S$ in Eq.(\ref{eq:gd}), with $\epsilon$ a small number. The calculation of $S$ can become the bottleneck in case the number of parameters is too large. This issue could be leveraged by representing $S$ as a matrix function instead of building it explicitly~\cite{NETKET3}, or by freezing a large portion of $S$ during each iteration similar to DMRG~\cite{ZhangPoletti2022}. Here this is not a significant concern, because we use at most about $1000$ parameters to specify the network.
To further enhance the stability of the algorithm, we add the contribution of an L2 regularization term when evaluating the gradient in Eq.(\ref{eq:gd}), that is, instead of directly choosing $F$ as the gradient of $\tilde{E}(\theta)$, $F$ is chosen as the gradient of the function $\tilde{E}(\theta) + \lambda ||\theta ||^2$ instead where $||\cdot||^2$ means the square of the Euclidean norm. In this work we choose $\epsilon=0.02$ and $\lambda=10^{-3}$ for our numerical simulations if not particularly specified.

\section{Results} \label{sec:result}

\subsection{Training Details}
In this work we use the Adam optimizer~\cite{Adam} for the VMC iterations, with an initial learning rate of $\alpha= 0.001$, and the decay rates for the first- and second-moment to be $\beta_1=0.9$, $\beta_2=0.99$ respectively.
For the Metropolis-Hastings sampling, we will use a fixed $N_s=4\times 10^4$ for our numerical simulations if not particularly specified (in principle one should use a larger $N_s$ for larger systems, however in this work we focus on molecular systems with at most $30$ qubits). We will also use a thermalization step of $N_{th} = 2\times 10^4$ (namely throwing away $N_{th}$ samples starting from the initial state). To avoid auto-correlation between successive samples we will only pick one out of every $10N_v$ samples. In addition, for each simulation we run $8$ Markov chains, and the energy is chosen to be the lowest of them. Since the energy will always contain some small fluctuations when $N_s$ is not large enough, the final energy is evaluated by averaging over the energies of the last $20$ VMC iterations.



\subsection{Effect of hidden size}
\begin{figure}
\centering
\includegraphics[width=0.9\columnwidth]{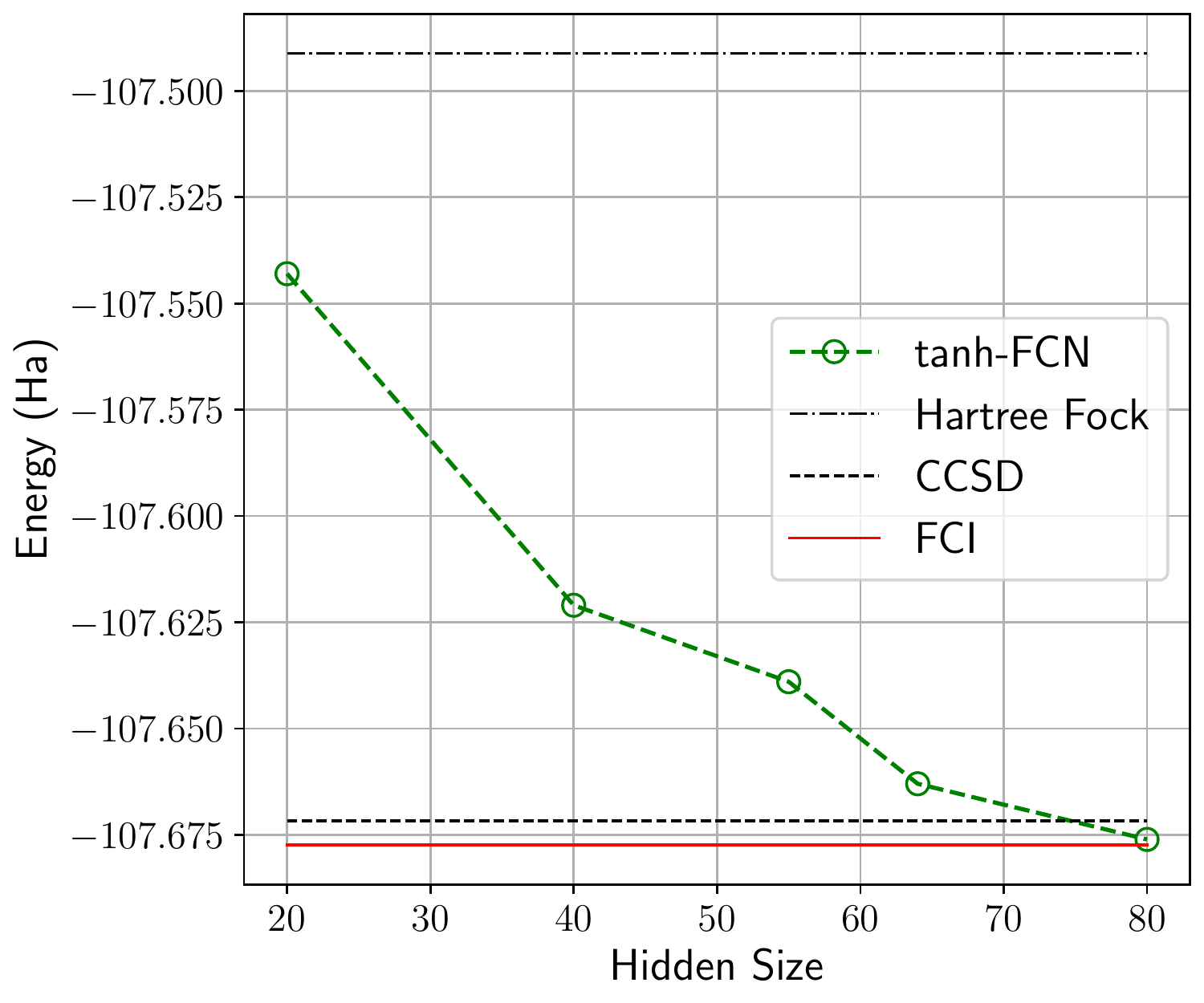}
\caption{Influence of the number of hidden spins in our tanh-FCN on the accuracy of the final energy. The N2 molecule in the STO-3G basis is used.}
\label{fig:n2-hidden-size}
\end{figure}

We first study the effect of $N_h$ which essentially determines the number of parameters, thus the expressivity of our tanh-FCN (analogously to RBM). The result is shown in Fig.~\ref{fig:n2-hidden-size} where we have taken the N2 molecule as an example. We can see that by enlarging $N_h$, the precision of tanh-FCN can be systematically improved. With $N_h =4N_v = 80$, we can already obtain a final energy that is lower than the CCSD results.


\subsection{Potential Energy Surfaces}

\begin{figure}
\centering
\includegraphics[width=1.0\columnwidth]{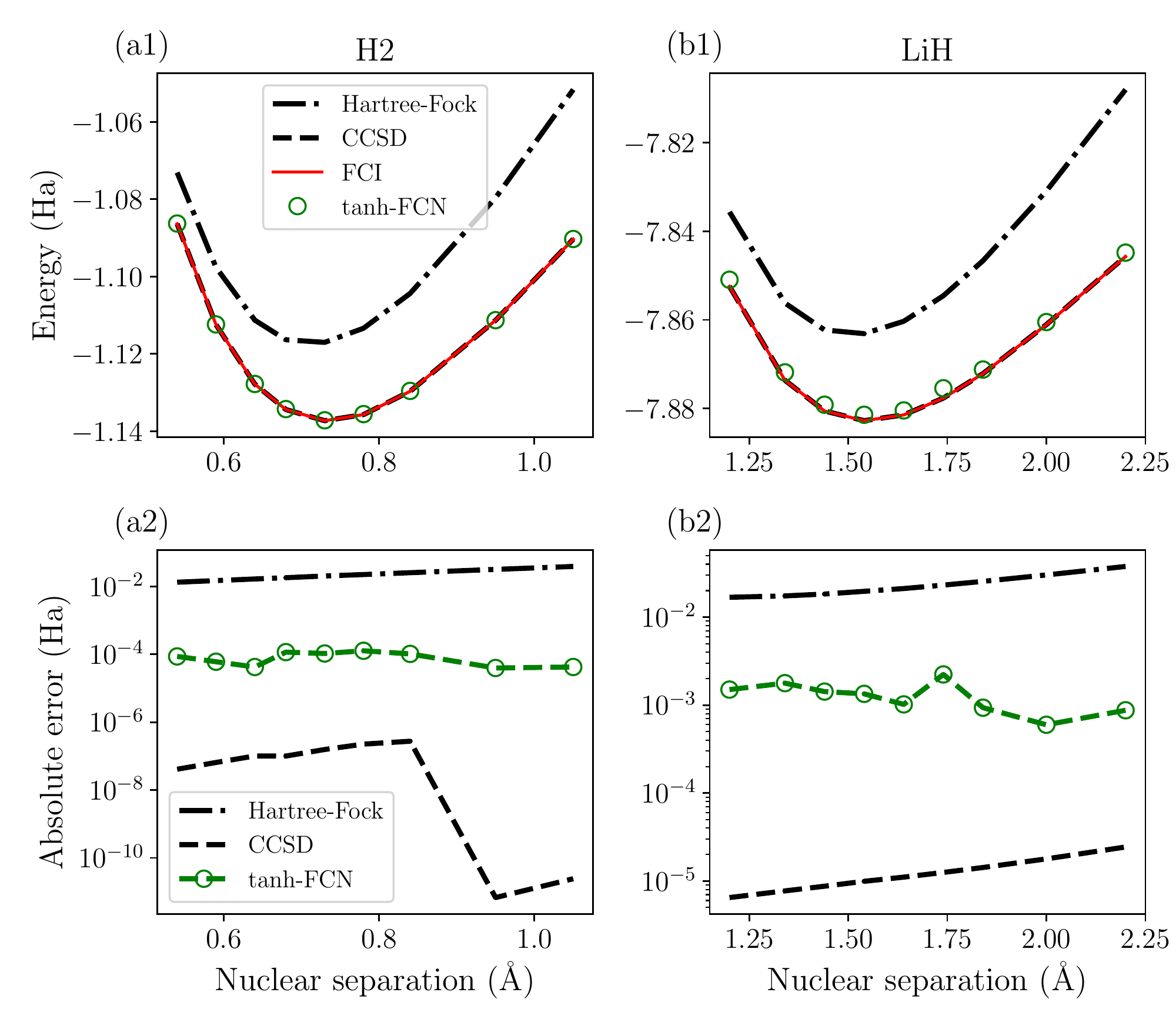}
\caption{Potential energy surfaces of (a1) H2 and (b1) LiH. We have used $N_h/N_v=2$ for H2 and $N_h/N_v=4$ for LiH, which are sufficient for our tanh-FCN to reach chemical precision. We have also used $N_s=2\times 10^4$ for both molecules during the training. (a2) and (b2) show the absolute error with respect to the FCI energy for H2 and LiH respectively. 
}
\label{fig:PES}
\end{figure}

Now we demonstrate the accuracy of our tanh-FCN by studying the potential energy surfaces of the two molecules H2 and LiH in the STO-3G basis, as shown in Fig.~\ref{fig:PES}. We can see that for both molecules under different bond lengths, our simulation can reach lower or very close to the chemical precision, namely error within $1.6\times 10^{-3}$ Hatree (Ha) or 1 kcal/mol (CCSD results are extremely accurate for these two molecules). 


\subsection{Final energies for several molecular systems}

\begin{table}[!ht]
    \begin{center}
    \caption{List of molecules and the ground state energies computed using RBM, tanh-FCN, CCSD. The FCI energy is also shown as a reference. The column $N_v$ shows the number of qubits. We have used $N_h/N_v=2$ for all the molecules studied.}
    \label{tab:ground-states-compare}
    \begin{tabular}{cccccc}
    \hline
    \hline
        \textbf{Molecule} & $N_v$  & \textbf{RBM}~\cite{ChooCarleo2020} & \textbf{tanh-FCN} & \textbf{CCSD} & \textbf{FCI}  \\
        $\textbf H_2$ & $4$ & $-1.1373$ & $-1.1373$ & $-1.1373$ & $-1.1373$ \\ 
        \textbf{Be} & $10$ & - & $-14.4033$ & $-14.4036$ & $-14.4036$ \\ 
        \textbf{C} & $10$ & -  & $-37.2184$ & $-37.1412$ & $-37.2187$ \\
        $\textbf {Li}_2$ & $20$ & -  & $-14.6641$ & $-14.6665$ & $-14.6666$ \\
        \textbf{LiH} & $12$ & $-7.8826$ & $-7.8816$ & $-7.8828$ & $-7.8828$ \\ 
        $\textbf {NH}_3$ & $16$ & $-55.5277$ & $-55.5101$ & $-55.5279$ & $-55.5282$ \\ 
        $\textbf {H}_2 \textbf O$ & $14$ & $-75.0232$ & $-75.0021$ & $-75.0231$ & $-75.0233$ \\ 
        $\textbf {C}_2$ & $20$ & $-74.6892$ & $-74.6134$ & $-74.6744$ & $-74.6908$ \\ 
        $\textbf {N}_2$ & $20$ & $-107.6767$ & $-107.622$ & $-107.6716$ & $-107.6774$ \\ 
        $\textbf {CO}_2$ & $30$ & - & $-185.1247$  & $-184.8927$ & $-185.2761$ \\ 
        \hline
    \end{tabular}
    \end{center}
\end{table}

We further compare the precision of tanh-FCN with RBM and CCSD for several small-scale molecules in STO-3G basis, which are shown in Table.~\ref{tab:ground-states-compare}. For these simulations we have used $N_h/N_v=2$, while the RBM results are taken from Ref.~\cite{ChooCarleo2020}. These results show that even with a relatively small number of parameters and a real neural network, we can still obtain the ground state energies of a wide variety of molecules to very high precision (close to or lower than the CCSD energies). In the meantime, we note that the energies obtained using tanh-FCN is not as accurate as those obtained using RBM, however the computational cost of tanh-FCN is at least two times lower than RBM under with the same $N_h$ and we could relatively easily study larger systems such as CO2 with $30$ qubits.

\subsection{Effect of Hartree-Fock re-initialization}

\begin{figure}
\centering
\includegraphics[width=1.0\columnwidth]{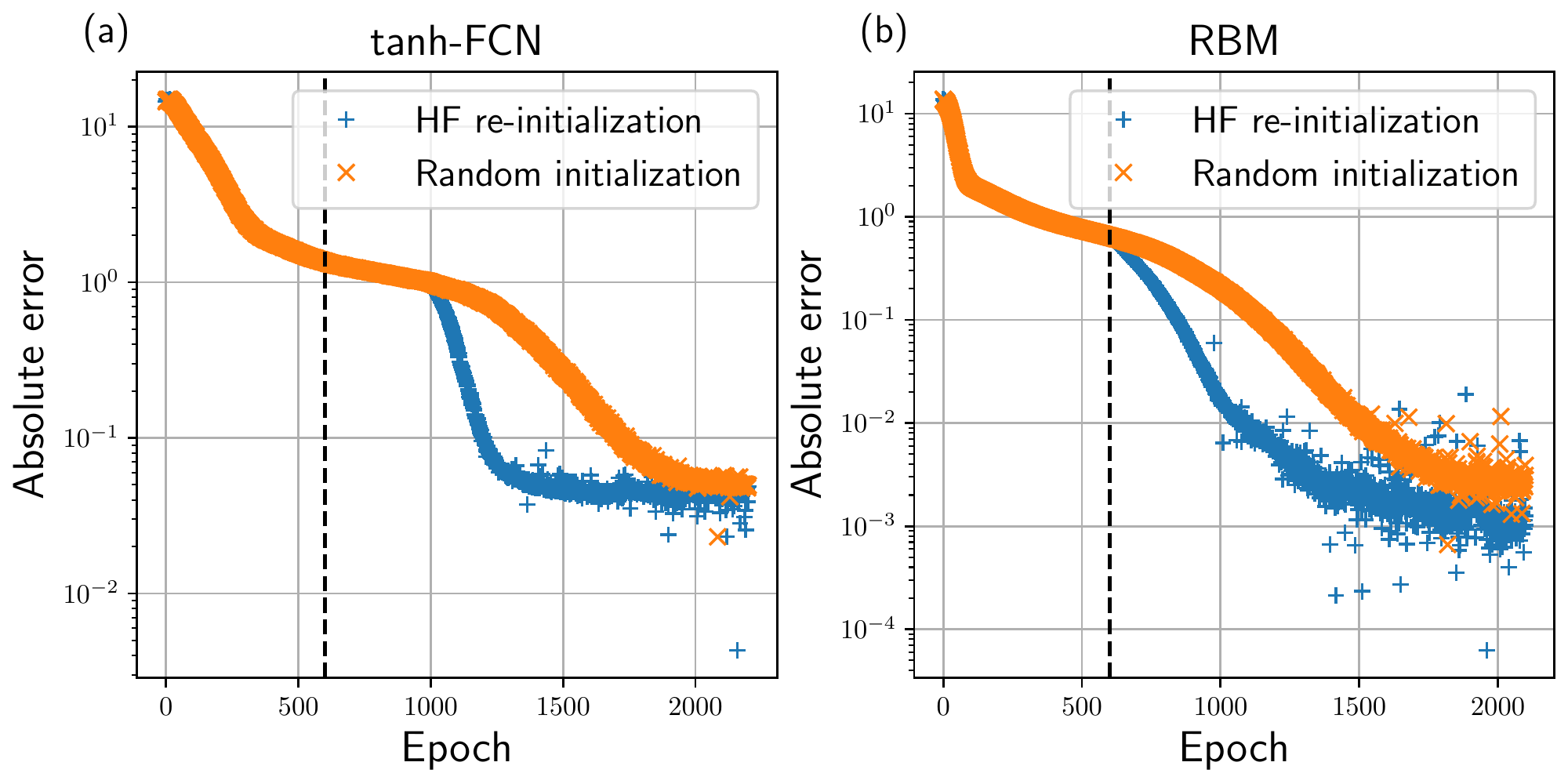}
\caption{Effect of the Hartree-Fock (HF) re-initialization compared to random initialization for (a) tanh-FCN and (b) RBM. The H2O (STO-3G basis, $14$ qubits) molecule is used here. The y-axis is the absolute error between the VMC energies and the FCI energy. For both methods we start to use the HF re-initialization starting from $600$-th VMC iteration marked by the vertical dashed lines. The other parameters used are $N_s=2\times 10^4$, $N_h/N_v=1$ and $\lambda=10^{-4}$.}
\label{fig:h2o-HFinit}
\end{figure}

There are generally two ingredients which would affect the effectiveness of the NNQS algorithm: 1) the expressivity of the underlying neural network ansatz and 2) the ability to quickly approach the desired parameter regime during the VMC iterations. The former is dependent on an intelligent choice of the neural network ansatz.
The effect of the latter is more significant for larger systems, and one generally needs to use a knowledged starting point such as transfer learning~\cite{ZenBressan2020,HebertBressan2020} for the VMC algorithm to guarantee success. For molecular systems it is difficult to explore transfer learning since the knowledge for different molecules can hardly be shared. However, for molecular systems the Hartree-Fock reference state may have a large overlap with the exact ground state, and is often used as a first approximation of the ground state. 
Here we show that for quantum chemistry problems the ability to reach faster the ground state can be improved by using the knowledge of the Hartree-Fock reference state. Concretely, during the VMC iterations, after the energies have become sufficiently close to the ground state energy, we stop using random initialization for our 
Metropolis-Hastings sampling, but use the Hartree-Fock reference state instead (Hartree-Fock re-initialization). The effect of the Hartree-Fock re-initialization is demonstrated in Fig.~\ref{fig:h2o-HFinit}, where we have taken the H2O molecule as our example. To show the versatility of the Hartree-Fock re-initialization, we demonstrate its effect for RBM as well. We can see that for both tanh-FCN and RBM, using Hartree-Fock re-initialization after a number of VMC iterations can greatly accelerate the convergence and reach a lower ground state energy than using random initialization throughout the VMC optimization. We can also see that for the H2O molecule tanh-FCN is less accurate than RBM using the same $N_h$, which is probably due to the fact that under the same $N_h$ tanh-FCN has a different expressive power as RBM for H2O.


\section{Conclusion}
\label{sec:summary}

We propose a fully connected neural network inspired from the restricted Boltzmann machine to solve quantum chemistry problems. Compared to RBM, our tanh-FCN is able to output both positive and negative numbers even if the parameters of the network are purely real. As a result we can directly study quantum chemistry problems using tanh-FCN with real numbers. In our numerical simulation, we demonstrate that tanh-FCN can be used to compute the ground states with high accuracy for a wide range of molecular systems with up to $30$ qubits. In addition, we propose to explicitly use the Hartree-Fock reference state as the initial state for the Markov chain sampling used during the VMC algorithm and demonstrate that this technique can significantly accelerate the convergence and improve the accuracy of the final energy for both tanh-FCN and RBM. Our method could be used in combination with existing high performance computing devices which are well optimized for real numbers, such as to provide a scalable solution for large-scale quantum chemistry problems.


\begin{acknowledgements}
We thank Xiao Liang, Mingfan Li for helpful discussions of the algorithm. 
C. G. acknowledges support from National Natural Science Foundation of China under Grant No. 11805279. H. S. acknowledges support from the National Natural Science Foundation of China (22003073, T2222026). D.P. acknowledges
support from the National Research Foundation, Singapore under its QEP2.0 programme (NRF2021-QEP2-02-
P03).

\end{acknowledgements}

\bibliographystyle{apsrev4-1}
\bibliography{qc}

\end{document}